# The Human Capital Ontology*

Shane Babcock[1,*], Maxwell Farrington[1,†], and John Gugliotti[1,†]

[1] *Knowledge and Decision Sciences (KadSci) LLC, 10300 Eaton Place, Suite 440, Fairfax VA, 22030*

**Abstract**

The Human Capital Ontology (HCO) is an ontology that represents data standards maintained and employed by the Office of Personnel Management (OPM) to represent Human Capital Operations and to classify job positions. The HCO is an extension of the Common Core Ontologies and the upper-level Basic Formal Ontology (BFO). HCO provides representation of OPM Natures of Action (NOA) that are used to identify human resource personnel actions, as well as their corresponding codes. HCO also represents Occupational Groups and Job Families, the Occupational Series into which these subdivide, as well as their corresponding codes, used by OPM to classify and grade both white- and blue-collar jobs in the Federal Government. HCO also encodes crosswalks between OPM Occupational Series and corresponding Standard Occupational Classification Codes maintained by the U.S. Bureau of Labor Statistics.

## 1. Introduction

The United States Office of Personnel Management (OPM) employs a variety of data standards with which to represent Human Capital Operations, and to classify and grade occupations, within the Federal Government. Under the statutory authority of the Classification Act of 1949, the OPM publishes and maintains classification standards programs for positions in the General Schedule [1]. These standards define various classes of positions in terms of their duties, responsibilities, and qualification requirements, establish official class titles, and set forth grades in which those classes have been placed. Federal agencies are to use these standards to place the organization's positions into their proper classes and grades.

While there exist similar efforts to represent the domain of occupational standards [2][3], to our knowledge there does not yet exist an ontology that models the occupational structure and classification system developed by the OPM for positions in the General Schedule. This paper describes the Human Capital Ontology (HCO), an ontology designed to fill this gap.

Note that we adopt the following typographical convention in this paper: ontology classes are expressed in small caps (e.g. POSITION, PERSONNEL ACTION) whereas ontology relations are expressed in bolded, italicized lowercase with words joined by underscores (e.g. ***has_part***, ***bearer_of***).

1* Corresponding author.
† These authors contributed equally.
sbabcock@kadsci.com (S. Babcock); mfarrington@kadsci.com (M. Farrington); jgugliotti@kadsci.com (J. Gugliotti)
0000-0003-0798-114X (S. Babcock); 0009-0006-3176-8823 (J. Gugliotti)

## 2. Methods

### 2.1. Imports

For HCO, we took as our starting point the top-level Basic Formal Ontology (BFO) [4] and the mid-level Common Core Ontology (CCO) suite [5]. The CCO provides classes representing different types of information content entities from which we extended to define key HCO classes such as POSITION, OCCUPATIONAL SERIES, NATURE OF ACTION IDENTIFIER, and NATURE OF ACTION CODE. Additionally, HCO imports the CCO Document Acts Ontology (CCO-D-Acts) which fills a gap in the representation of social, legal, normative, and deontic entities for ontologies that rely on the CCO suite [6]. The CCO-D-Acts representation of deontic roles, in particular duty holder roles, are used to connect positions to corresponding roles that persons bear when they are employed in a given position. Data used to build the HCO model was gathered from documentation published online by the OPM, as well as related portions of United States Code and the Code of Federal Regulations [1, 7, 8].

### 2.2. Data Retrieval

The data used to build the HCO was contained within a series of public documents published by the Office of Personnel Management. To access and utilize this data, the normal use case would require extensive searching and switching between documents to see the full picture of the hierarchy that is created by this series of documents. In building our ontology, this process was automated using a combination of open-source python libraries to convert the PDF documents to textual data and re-trieve the desired data to then convert into our ontology. The PDF documents were processed into textual data using pypdfium2 [9]. This textual data was then parsed using regular expressions to gather the codes, titles, and descriptions of different job families, job groups, occupational series, and nature of action codes. Due to the inherent structure of this data, much of the ontology building process was then able to be automated using RDFLib [10].

## 3. Results

### 3.1. Positions and Work Duties

The OPM classification system starts with the notion of a position. The OPM adopts the definition of position provided in 5 U.S. Code § 5102: 'position means the work, consisting of the duties and responsibilities, assignable to an employee.' Notably, organizational positions are often described as being vacant, in which case an organization publishes a job announcement seeking applicants to fill the position. A position can thus be thought of as an officially authorized organizational personnel requirement to perform certain duties and responsibilities that may or may not be filled at any given time. When a person becomes employed under a given position title, that person takes on the relevant duties and responsibilities as their own, thus filling the personnel requirement.

This raises the issue of how to categorize the class POSITION. One might think that a position is a kind of BFO ROLE, such as the teacher role, or the student role. But one cannot represent POSITION as a subclass of ROLE. ROLE is a subclass of BFO's SPECIFICALLY DEPENDENT CONTINUANT (SDC). A SDC is an entity that depends for its existence upon some specific entity that is its bearer. The color of my hair, the shape of my head, my weight, and my height are all examples of instances of SDC. The existence of each of these SDCs depends upon me, or one of my body parts, as its bearer. So too for roles. A given instance of teacher role or student role always depends upon the existence of the person who is its bearer.

This is not the case with positions, although there is certainly an important link between positions and roles. Suppose a company is advertising a job opening for a newly created position of Senior

Ontologist. The position exists unfilled until someone is interviewed and hired to fill that position. At that point, the person hired to fill the position becomes bearer of an instance of Senior Ontologist role. If that person is later separated from the company, that person ceases to be the bearer of that instance of the role, in which case that instance of the role ceases to exist. The position, on the other hand, may remain as a vacancy to be refilled (assuming that the company doesn't decide to eliminate the position itself). Thus, the position is one thing, the role that a person gains upon accepting the position and loses upon being separated from the company, is another.

While an instance of the class POSITION does not depend for its existence upon a specific bearer in the way that an instance of ROLE does, there is a dependency at play. In our previous example, suppose that after separating from the initial position holder the company later hires a new person to fill its Senior Ontologist position. During two distinct periods of time, two different people will have held or occupied the same instance of Senior Ontologist position. In HCO, we classify POSITION as a subclass of BFO's GENERICALLY DEPENDENT CONTINUANT (GDC). A GDC is an entity that can be characterized as a content or pattern that can be shared by multiple copies. A paradigm example of a GDC is a pdf file. Copies of the very same pdf file can be saved on multiple different physical storage devices (e.g. on the hard drives of various computers). Thus, that pdf file is always dependent upon *some* physical bearer or another. But its existence does not specifically depend upon any one of those bearers in particular [11]. An instance of GDC is related to its bearers via the BFO relation ***generically_depends_on***. An instance of GDC *b* is said to generically depend on another entity *c* if *c* is the bearer of an instance of SDC that ***concretizes*** *b*. For example, the Coca Cola trademark is a GDC that generically depends on each of the labels of the Coco Cola bottles in a convenience store cooler. Each bottle's label is the bearer of a distinct SDC, a quality pattern of red and white swirls, that concretizes the same Coca Cola trademark [11]. Our contention is that a position is a generically dependent entity in this sense.

In HCO we categorize POSITION as a subclass of the CCO class INFORMATION CONTENT ENTITY. An INFORMATION CONTENT ENTITY is a generically dependent continuant that depends on some information bearing entity and is about some entity. Examples include the content of a news report describing an event, a person's name, or the content of photograph which represents the pictured entity. More specifically, we assert POSITION to be a subclass of CCO's DIRECTIVE INFORMATION CONTENT ENTITY (DICE). A DICE is an INFORMATION CONTENT ENTITY that prescribes something. Examples include rules or regulations which prescribe that people ought to or must act in a certain way, or a blueprint design for a car. Building on the 5 U.S. Code § 5102 definition of a position as consisting of the duties and responsibilities assignable to an employee, we define POSITION as follows: A DIRECTIVE INFORMATION CONTENT ENTITY that consists of a set of responsibilities and duties an AGENT is expected to perform during some ACT OF EMPLOYMENT, and which is assigned to that AGENT by a competent authority. The idea is that a position is a set of prescriptions to perform certain types of work for a company. When a person fills a position, they take on the relevant duties and responsibilities as their own.

We also introduce a further subclass of DICE, the class WORK DUTY. We define WORK DUTY as a part of some POSITION which requires an AGENT to perform some work activity during some ACT OF EMPLOYMENT in that POSITION. For example, an ontologist position has as part a work duty that prescribes that any person holding the position should perform work developing ontologies. We also introduce a subclass of WORK DUTY, namely PRIMARY DUTY. Building on Code of Federal Regulations definitions of the term primary duty [13, 14], we define the class PRIMARY DUTY as follows: A WORK DUTY that constitutes a basic reason for the existence of the POSITION of which it is a part, and which requires that some work activity occupy a substantial portion of an AGENT's time over a typical work cycle and that it be assigned on a regular and recurring basis. To return to the previous example, the development of ontologies is a primary duty of the ontologist position, as it is a requirement that this work should occupy a substantial portion of an ontologist's time and that it should be assigned on a regular and recurring basis.

When a person is hired to fill an instance of POSITION, they become the bearer of a corresponding instance of ROLE. To represent the connection between positions and roles, we introduced a new relation to HCO, ***holds_position*.** This relation is a subproperty of BFO's ***is_carrier_of*** relation, itself the inverse of ***generically_depends_on***. We define the ***holds_position*** relation as follows: x holds position y iff y is an instance of POSITION and x is an instance of AGENT, such that each WORK DUTY that is part of y is concretized by some DUTY HOLDER ROLE inhering in x. (The class DUTY HOLDER ROLE is imported from CCO-D-Acts). Elucidation: a position consists of a set of duties and responsibilities to perform certain work, and thus when a person is appointed to fill a position, they thereby accept those duties and responsibilities as their own. In doing so, the person becomes the bearer of various duty roles corresponding to the duties of the position, duty roles which if realized, are realized in the performance of the work prescribed by the position's work duties.

The HCO treatment of a POSITION as a DICE consisting of a set of work duties and responsibilities allows us to explain why positions can often exist vacant or unfilled. Figure 1 provides an illustration of how an instance of POSITION can exist during one temporal interval independently of any person bearing a corresponding role and then later during another interval come to generically depend upon a person bearing such a role.

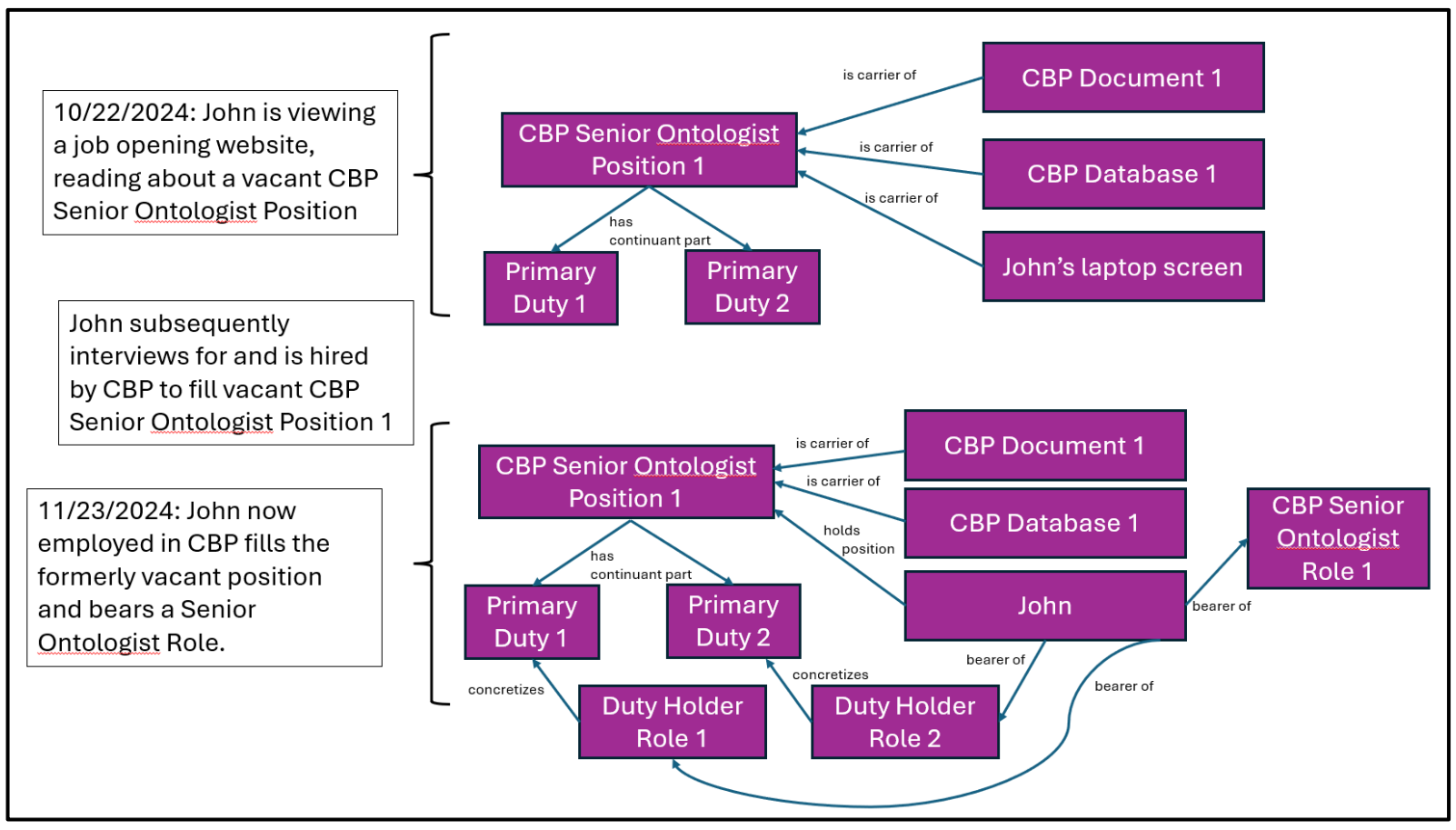

**Figure 1:** A CBP Senior Ontologist Position changing from an unfilled position to a filled position.

The same information content can be carried by multiple information bearers, including the work prescriptions that are constitutive of a position. Thus, suppose that on 10/22/2024 John is sitting at his laptop viewing a job opening website, reading about a vacant instance of the position: Customs and Border Protection (CBP) Senior Ontologist Position (Figure 1). John does some research on the work the position involves by reviewing its constitutive work duties, including primary duties (as illustrated above, an instance of POSITION can be linked to the various instances of PRIMARY DUTY that it involves by means of the BFO relation ***has_continuant_part***.) The screen on which John reads about the position's work duties is one carrier of that information. But that information also generically depends upon other information bearing entities, including an instance of a paper CBP document and an instance of a CBP database that both contain various official CBP position descriptions. Subsequently, John interviews for and is hired to fill CBP's Senior Ontologist Position. Newly employed by CBP, on 11/23/2024 John now holds the Senior Ontologist Position. John (more specifically his brain) is also now a carrier of that information in addition to the CBP position description document and database. In John's case, he carries the position by virtue of bearing instances of DUTY HOLDER ROLE which concretize the position's various constitutive work duties (including primary duties). If John is later removed from that position (or leaves for another job), then assuming that CBP retains the position, it will once again exist as a vacant position until someone else is hired to fill it. During that time no one will *hold* the position, but there will still exist carriers of the information (such as the CBP document and database).[2]

### 3.2. Occupational Groups, Job Families, and Occupational Series

The OPM classifies Federal Government positions, both white- and blue-collar, into occupational groups or job families. An occupational group is a major subdivision of the white-collar occupations covered by the General Schedule, drawn by the OPM, that embraces a group of associated or related occupations. Whereas a job family is a broad group of trades, craft, and labor occupations related in one or more ways such as: similarity of functions performed, transferability of knowledge and skills from one occupation to another, or similarity of materials or equipment worked on. [1, 7]. An example of an occupational group is the *Inspection, Investigation, Enforcement, and Compliance Group* which is a broad grouping of positions the duties of which are to advise on, administer, supervise, or perform inspection, investigation, enforcement, or compliance work primarily concerned with alleged or suspected offenses against the laws of the United States, or work primarily concerned with determining compliance with laws and regulations. An example of a job family is the *Machine Tool Work Family*, which consists of those jobs that involve setting up and operating machine tools and using hand tools to make or repair metal parts, other tools, gauges, models, patterns, mechanisms, and machines, as well as the machining of explosives and synthetic materials. The OPM further subdivides each occupational group and job family into numerous distinct subgroups of positions called *occupational series*. An occupational series is a subdivision of an occupational group or job family that consists of positions that are similar with respect to specialized line of work and qualification requirements [1, 7]. Thus, for example, the *Inspection, Investigation, Enforcement, and Compliance Group* is subdivided into occupational series including the *Border Patrol Enforcement Series,* the *Customs and Border Protection Series*, the *Aviation Safety Series*, the *Food Inspection Series*, and the *General Investigation Series* among others.

Importantly, occupational groups, job families, and the series into which they are subdivided, are all described by the OPM as *groups* of positions. The *Border Patrol Enforcement Series*, for example, is a group of positions that includes, among others, *border patrol agent, supervisory border patrol agent*, *patrol agent in charge*, *chief patrol agent*, *watch commander*, and *special operations supervisor*. Of course, one can also represent those position types as subclasses of a broader class such as UNITED STATES BORDER PATROL POSITION (we do so elsewhere in work that extends from the HCO), itself an indirect subclass of POSITION. But a group of positions, which is what the OPM has in mind, is not a type of position, and so it would be a mistake to represent an occupational series as a subtype of POSITION.

Since we have defined POSITION as an INFORMATION CONTENT ENTITY, it made sense to also represent a group of positions as some sort of INFORMATION CONTENT ENTITY (an aggregate of information is itself information). In order to represent occupational groups, job families, and occupational series, the HCO therefore introduced the class AGGREGATE OF POSITIONS defined as an INFORMATION CONTENT ENTITY that has some POSITIONS as parts. We define the class OCCUPATIONAL GROUP as an AGGREGATE OF POSITIONS that is a major subdivision of the White Collar Occupations covered by the General Schedule, embracing a group of associated or related occupations. We define the class TRADE, CRAFT, OR LABOR JOB FAMILY as an AGGREGATE OF POSITIONS consisting of a broad grouping of trades, craft, and labor occupations related in one or more ways such as: similarity of functions performed, transferability of knowledge and skills from one occupation to another, or similarity of materials or equipment worked on. And we define the class Occupational Series as an Aggregate of Positions that is part of an Occupational Group or Job Family and which has as parts Positions that are similar as to specialized line of work and qualification requirements. Specific occupational groups, job families, and occupational series are then introduced in HCO as instances of the classes Occupational Group, Trade, Craft, or Labor Job Family, and Occupational Series. Each instance of Occupational Group and Trade, Craft, or Labor Job Family is provided with object property assertions involving the BFO relation ***has_continuant_part*** that link that group or family to the various instances of occupational series into which it subdivides. Figure 2, for example, illustrates the links between the *Inspection, Investigation, Enforcement, and Compliance Group* and the instances of Occupational Series into which it subdivides, including the *Border Patrol Enforcement Series*.

[2] Note while John no longer *holds* the position in that case, he may still be a *carrier* of the information. After all, unless John suffers serious memory loss, he will still retain a knowledge of the position's duties and responsibilities within his brain. The difference is that John no longer recognizes those duties and responsibilities as his own—his brain ceases to concretize them as instances of duty holder roles.

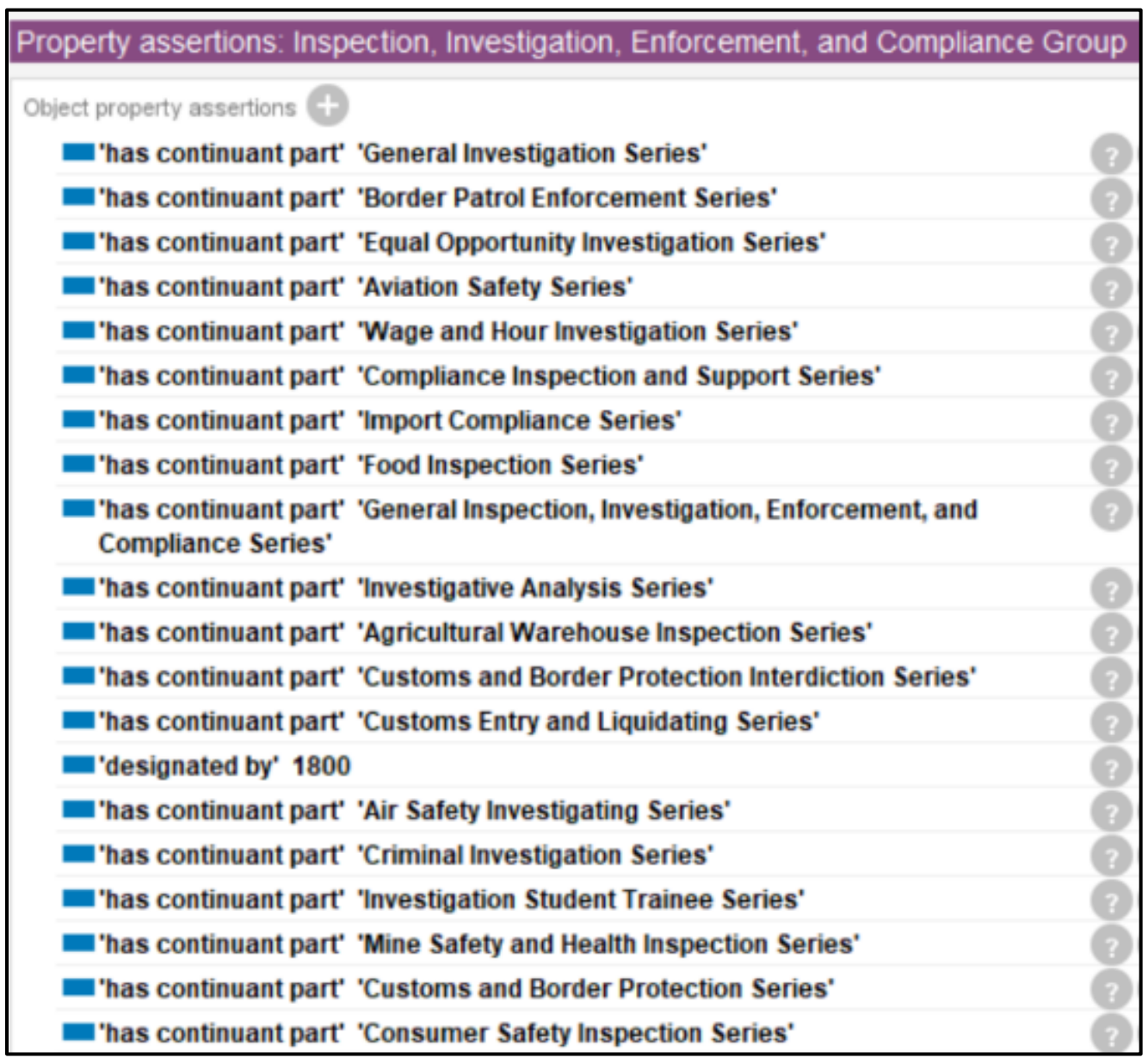

**Figure 2:** Occupational Series that are part of the Inspection, Investigation, Enforcement, and Compliance Group

For each instance of Occupational Group, Trade, Craft, or Labor Job Family, and Occupational Series, the OPM provides a 4-number code that uniquely identifies it. We represent each of these codes in HCO as instances of the CCO class Code Identifier. A Code Identifier is an information content entity that is used to designate some entity, and which consists of a string of characters that was created and assigned according to an encoding system such that metadata can be derived from the identifier. We introduce the classes Occupational Group Code and Job Series Code in HCO as subclasses of Code Identifier. The former is defined as a Code Identifier that designates some instance of Occupational Group or Trade, Craft, or Labor Job Family, whereas the latter is defined as a Code Identifier that designates some instance of Occupational Series. Specific occupational group codes and job series codes provided by the OPM are represented in HCO as individuals, each of which is asserted as an instance of either the class Occupational Group Code or Job Series Code. In HCO, instances of these codes are linked to corresponding instances of occupational group, job family, and occupational series by the CCO ***designates*** relation and its inverse the ***designated_by*** relation. Thus, as illustrated in Figure 2, the *Inspection, Investigation, Enforcement, and Compliance Group* is ***designated_by*** the occupational group code *1800*. Conversely, the *1800* code ***designates*** the *Inspection, Investigation, Enforcement, and Compliance Group*. Similarly, the job series code *1895* ***designates*** the *Customs and Border Protection Series*, the job series code *1896* ***designates*** the *Border Patrol Enforcement Series*, and so on.

### 3.3. OPM Natures of Action and Nature of Action Codes

By law, the Office of Personnel Management has the authority to prescribe reporting requirements for personnel actions, such as appointments and separations, taken by Government Executive Agencies [12]. To document personnel actions, Agencies make use of the Standard Form 50, Notification of Personnel Action.[3] The form includes a variety of data fields, one of the most important of which has the header 'Nature of Action'. A nature of action (NOA) is a standardized phrase used by OPM, such 'Appointment Status Quo', that is used to explain the personnel action being taken [8]. In other words, each NOA is a phrase that is used to refer to instances of personnel actions of different types when documenting those actions. For instance, consider personnel actions that assign a person to a competitive service position within an Agency (competitive service positions are those subject to the competitive examination requirements of the Civil Services Act). When a person is assigned to such a position under a temporary appointment, the personnel action is identified using the NOA 'Appointment Not to Exceed (Date)'—where the date specifies a calendar day which the appointment is not to exceed. Whereas, if a person is assigned to such a position under a *career appointment*, the personnel action is identified using the NOA 'Career Appointment.' And in the case where a person is assigned to a senior executive service position under a career appointment, the personnel action is identified using the NOA 'Senior Executive Service Career Appointment'. To model NOAs, we introduced the class Nature of Action Identifier as a subclass of CCO's Designative Information Content Entity, the latter of which is an Information Content Entity consisting of a set of symbols that denote some entity. We define the class Nature of Action Identifier as follows: A Designative Information Content Entity that consists of a phrase that designates some Personnel Action. Using [8] as our primary source, we modelled various personnel actions as subclasses of CCO's Planned Act. A snapshot of HCO's taxonomy of personnel actions is provided in Figure 3.

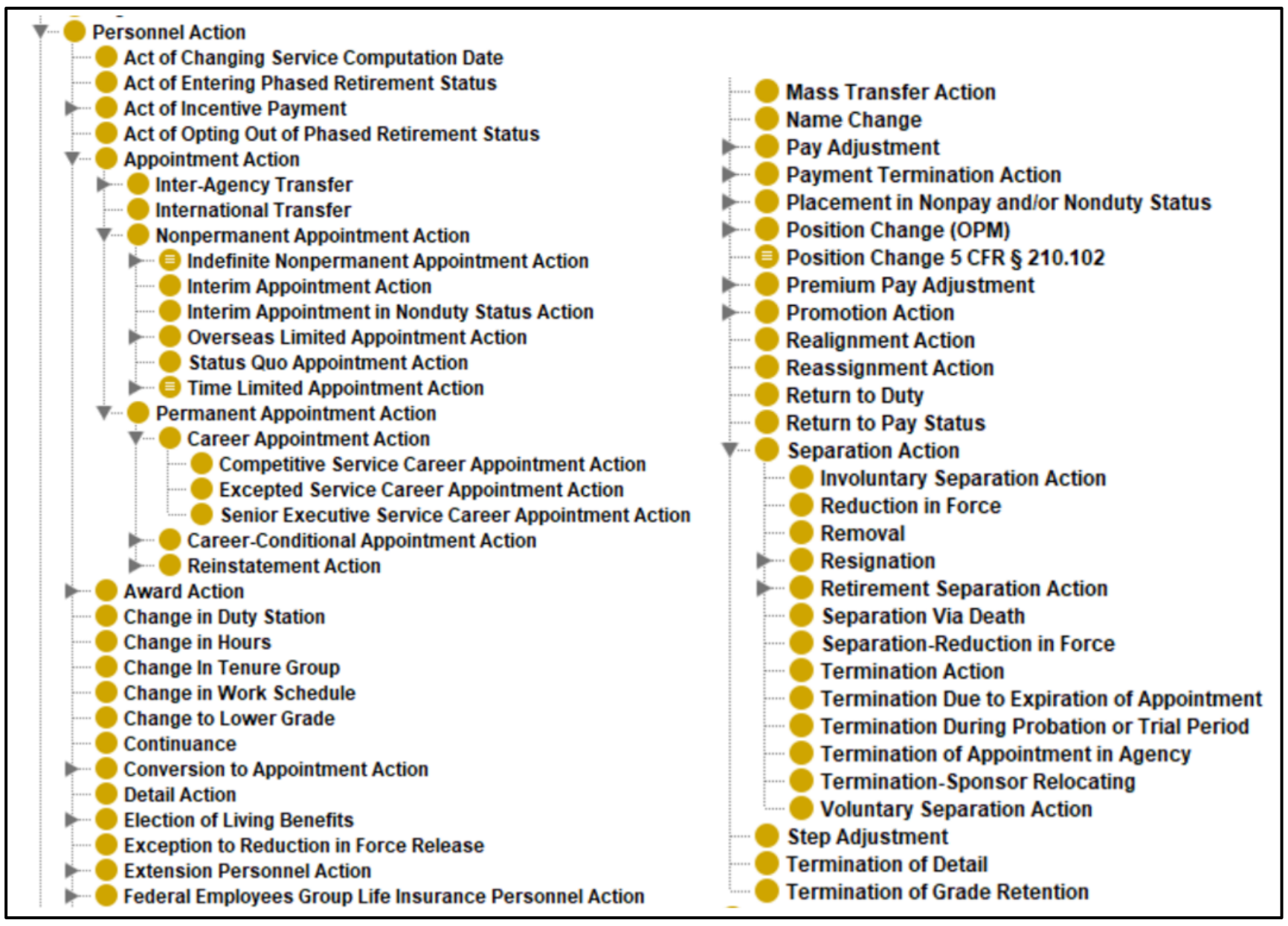

**Figure 3:** HCO's Taxonomical Representation of OPM Personnel Actions

For each NOA, the OPM provides a corresponding nature of action code that is to be entered into Standard Form 50 along with the NOA. These codes are used by the OPM to identify particular NOAs for statistical and data processing purposes [8]. Each of these codes is a numerical string consisting of 3 digits, ranging from 000 to 999. The codes are classified into series, such as the 100 series, 200 series, 300 series, and so on. The first digit of each code indicates the broad type of the personnel action designated by the corresponding NOA. In the case of the 100 series codes (100-199), the first digit '1' indicates that the code is an identifier for an NOA designating an appointment action. Whereas in the case of the 300 series codes, the first digit '3' indicates that code is an identifier for an NOA designating a separation action (a personnel action that ends a person's employment with an Agency). To model NOA codes we introduced the class NATURE OF ACTION CODE as a subclass of CCO's CODE IDENTIFIER. We define NATURE OF ACTION CODE as follows: A CODE IDENTIFIER that consists of a unique 3-digit numerical string that designates some NATURE OF ACTION IDENTIFIER for statistical and data processing purposes.

[3] A copy of the form can be found here: https://www.opm.gov/forms/pdfimage/sf50.pdf

Both the classes NATURE OF ACTION IDENTIFIER and NATURE OF ACTION CODE are asserted to have several subclasses in HCO. Subclasses of NATURE OF ACTION IDENTIFIER include, for example: APPOINTMENT NATURE OF ACTION IDENTIFIER defined as a NATURE OF ACTION IDENTIFIER that designates some APPOINTMENT ACTION; CONVERSION TO APPOINTMENT IDENTIFIER defined as a NATURE OF ACTION IDENTIFIER that designates some CONVERSION TO APPOINTMENT ACTION; and SEPARATION IDENTIFIER defined as a NATURE OF ACTION IDENTIFIER that designates some SEPARATION ACTION. Corresponding to these are the following subclasses of NATURE OF ACTION CODE: 100 SERIES NATURE OF ACTION CODE defined as a NATURE OF ACTION CODE which designates some APPOINTMENT NATURE OF ACTION IDENTIFIER; 300 SERIES NATURE OF ACTION CODE defined as a NATURE OF ACTION CODE which designates some SEPARATION NATURE OF ACTION IDENTIFIER; and 500 SERIES NATURE OF ACTION CODE defined as a NATURE OF ACTION CODE which designates some CONVERSION TO APPOINTMENT NATURE OF ACTION IDENTIFIER. Note that specific NOAs and NOA codes employed by the OPM are represented in HCO as *individuals*, not *classes*. For example, the *Career Appointment Identifier* and the *Career-Conditional Appointment Identifier* are both individual instances of the class APPOINTMENT NATURE OF ACTION IDENTIFIER. These identifiers are designated, respectively, by *Nature of Action Code 100* and *Nature of Action Code 101* both of which are individual instances of the class 100 SERIES NATURE OF ACTION CODE. The HCO model of the relationship between personnel actions, NOAs and their NOA codes is illustrated in Figure 4 (yellow rectangles represent classes whereas purpose rectangles represent instances of classes).

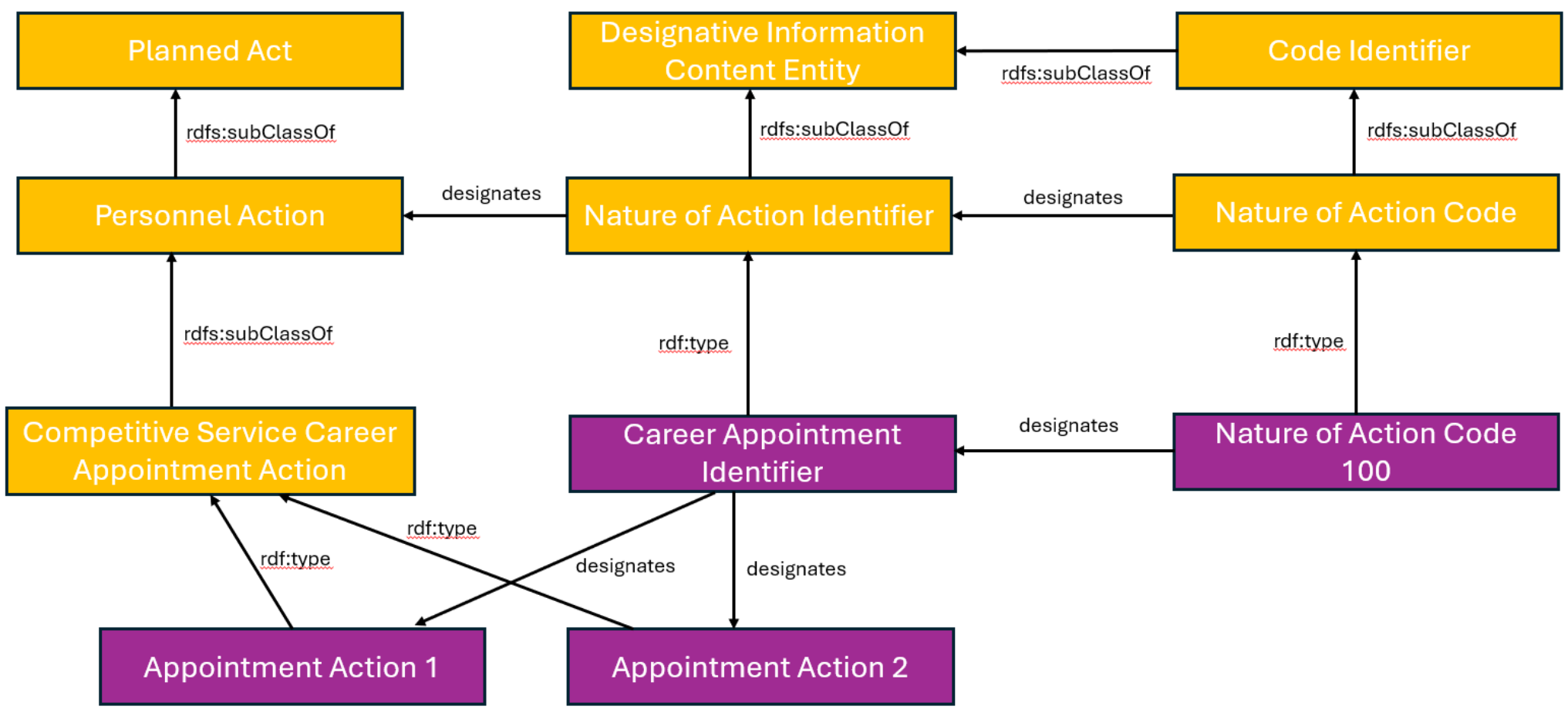

**Figure 4:** Relationships between Personnel Actions, Natures of Action, and Nature of Action Codes

Thus, suppose that two employees of a certain Agency are both appointed to competitive service positions under a career appointment. These appointment actions are represented in Figure 4 by *Appointment Action 1* and *Appointment Action 2*, both of which are instances of the class COMPETITIVE SERVICE CAREER APPOINTMENT ACTION. An official of the Agency will need to document both of these personnel actions on a separate copy of the Standard Form 50. On both copies of the form the official writes the phrase 'Career Appointment' in the nature of action data field. As a result each form contains a distinct bearer of the same information content, namely the *Career Appointment Identifier*.

In this way, the *Career Appointment Identifier* is used in two separate contexts to designate both *Appointment Action 1* and *Appointment Action 2*. The official also enters the string '100' into the nature of action code field on each form and in doing so produces two separate bearers of *Nature of Action Code 100*. In this way, *Nature of Action Code 100* is used in two separate contexts to designate the *Career Appointment Identifier*.

## 4. Discussion

HCO and ontologies that extend therefrom can be used define various subclasses of POSITION. Instances of those position types can be linked via BFO's ***continuant_part_of*** relation to the corresponding instance of OCCUPATIONAL SERIES. Thus HCO or extension ontologies have the potential to be used by government agencies to generate knowledge graphs modeling these linkages. As we develop HCO's representation of OCCUPATIONAL SERIES with further axiomatization, we envision HCO serving as a queryable data source for agencies to use in properly classifying organizational positions according to the OPM standards.

## Declaration on Generative AI

The authors have not employed any Generative AI tools.

## Acknowledgements

We would like to acknowledge the following for their valuable discussion which helped to improve the development of the Human Capital Ontology: Daniel Maxwell, John Beverley, Ryan Riccucci, Mark Jensen, and Barry Smith. We would also like to acknowledge two anonymous reviewers for their very helpful comments on this paper.

## References

[1] Office of Personnel Management, Introduction to position classification standards, 2009. URL: https://www.opm.gov/policy-data-oversight/classification-qualifications/classifying-general-schedule-positions/positionclassificationintro.pdf.

[2] J. Beverley, S. Smith, M. A. Diller, et. al., The occupation ontology (OccO): Building a bridge between global occupational standards, in: Proceedings International Workshop on Ontologies for Services and Society, Sherbrooke Canada, 2023. URL: https://ceur-ws.org/Vol-3637/paper13.pdf.

[3] Occupation Ontology. GitHub Repository. URL: https://github.com/Occupation-Ontology/ OccO.

[4] BFO Ontology. GitHub Repository. URL: https://github.com/BFO-ontology.

[5] CommonCoreOntologies. GitHub Repository. URL: https://github.com/CommonCoreOntology/CommonCoreOntologies.

[6] CCO Document Acts Ontology. Github Repository. URL: https://github.com/jonathanvajda/cco-d-acts.

[7] Office of Personnel and Management, Handbook of occupational groups and families, 2018. URL: https://www.opm.gov/policy-data-oversight/classification-qualifications/classifying-general-schedule-positions/occupationalhandbook.pdf.

[8] Office of Personnel and Management, The guide to processing personnel actions, 2017. URL: https://www.opm.gov/policy-data-oversight/data-analysis-documentation/personnel-documentation/#url=Personnel-Actions.

[9] Pypdifium2. GitHub Repository. URL: https://github.com/pypdfium2-team/pypdfium2/tree/stable.

[10] RDFLib. GitHub Repository. URL: https://github.com/RDFLib/rdflib.
[11] R. Arp, B. Smith, and A.D. Spear, Building Ontologies with Basic Formal Ontology, The MIT Press, Cambridge, Massachusetts, 2015.
[12] 5 CFR Part 9. URL: https://www.ecfr.gov/current/title-5/part-9.
[13] 5 CFR 831.902. URL: https://www.ecfr.gov/current/ title-5/section-831.802.
[14] 5 CFR 842.802. URL: https://www.ecfr.gov/current/ title-5/section-842.802.